**Interacting humans use forces in specific frequencies to exchange information by touch**

C. Colomer[1]*, M. Dhamala[2], G. Ganesh[3], J. Lagarde[1]

[1] EuroMov Digital Health in Motion, Univ Montpellier, IMT Mines Ales, Montpellier, France
[2] Department of Physics and Astronomy, Neuroscience Institute, Georgia State University, USA
[3] Laboratoire d'Informatique, de Robotique et de Microélectronique de Montpellier (LIRMM), Univ. Montpellier, CNRS, France
*Corresponding author's email: clementine.colomer@umontpellier.fr

**Abstract**
Object-mediated joint action is believed to be enabled by implicit information exchange between interacting individuals using subtle haptic signals within their interaction forces. The characteristics of these haptic signals have, however, remained unclear. Here we analyzed the interaction forces during an empirical dyadic interaction task using Granger-Geweke causality analysis, which allowed us to quantify the causal influence of each individual's forces on their partner's. We observed that the inter-partner influence was not the same at every frequency. Specifically, in the frequency band of [2.15-7] Hz, we observed inter-partner differences of causal influence that were invariant of the movement frequencies in the task and present only when information exchange was indispensable for task performance. Moreover, the inter-partner difference in this frequency band was observed to be correlated with the task performance by the dyad. Our results suggest that forces in the [2.15-7] Hz band constitute task related information exchange between individuals during physical interactions.

**Introduction**
Physical interaction is a fundamental mode of communication in human society. Physical interactions help us learn skills, like learning to walk from our parents (1, 2). It helps us coordinate and work together in daily life, like lifting a table together (3) or dancing with a partner (4). It lets us welcome and comfort others, and we use physical interactions to assist others, like helping an elderly person to stand up, or helping a rehabilitation patient to move. In everyday life, interpersonal interaction is mediated by social cues that can differ heavily with cultures. Thus, the person making a touch and the place on the body where it is made can lead to very different inferences (5). The mechanical information (force, vibration, pressure), during these touches are transmitted by somesthetic and kinesthetic receptors present all over our body (6, for review). It is known that during object-mediated physical interactions between two individuals, humans are able to estimate various features of their partner's behavior, like its similarity to one's own behavior (7) and movement goal (8,9), and predict the future actions of a partner (10-12). These estimations are likely used for role distribution (13-15) and to coordinate actions. It is generally accepted that these estimations and resulting coordination are enabled by the exchange of task related information through a so called "haptic channel" (16), which implicitly *influence* the behaviors of the interacting individuals (8, 9, 17, 18). However, it remains unclear what this haptic channel consists of and what specific features of the force signals contribute to this information exchange.

It is often considered that interacting pairs of individuals, or 'dyads', have an asymmetry in their roles, one being more as a leader and the other more as a follower (19). Furthermore, previous studies have suggested reactive forces to play a key part in physical coordination (7, 8). These studies motivated us to analyze the causal symmetry (20, 21) in the interaction forces using Granger causality (22) for possible signatures of information exchange between the interacting individuals. The method provides us with Granger causality values that give a quantitative estimate of the influence in each direction between the interacting partners.

Granger causality analysis uses Norbert Wiener's prediction theory (23) to quantify the degree to which the past of one process *A* influences the present and future of another process *B* and enables quantitative estimation of two distinct directions of influence, from A to B and from B to A. In the case of object-mediated dyadic physical interactions, the interaction forces generated by the two partners are arguably a key feature embodying the information exchange. In this study, we therefore analyzed the time series of interaction forces using the spectral Granger causality commonly known as Granger-Geweke causality (GGC) (24), in which Granger causality is derived in the frequency domain.

This framework, including a conceptual definition of causal influence and its application for data analysis, has been applied to brain imaging, as neural networks oscillate at discrete frequency bands (25). Distinct frequencies can also be found when executing movements. Dating back from Woodworth (1899) (26), the study of voluntary movement was influenced by the hypothesis of movement generation based on a sequence of submovements, motivating the search for oscillatory components (27, see 28 in the case of slow movement). A frequency analysis of movements found an oscillatory or intermittent component when tracking visually a target on a screen (29), or another actor's movement (30). The relation between balance and a fingertip physical light contact also occurred at characteristic frequencies (31, 32). However, to our knowledge, an oscillatory component to genuine physical interactions among humans is still unknown.

We developed an empirical object-mediated physical interaction task for human dyads and measured the forces applied by each individual while performing the task. The task required them to perform repetitive movements together, requiring both temporal (audio metronome) and spatial (visual) feedback for completion. We provided only one type of feedback to each interacting individual, thus making information exchange necessary to complete the task (Experiment 1), and compared it with the case when both had sufficient feedback to perform the task alone (Experiment 2). We hypothesized that there would be higher GGC values in the first experiment compared to the second, thus reflecting the information flow between subjects. We also hypothesized the presence of at least, two bands in our frequency analysis. The first encompassing the oscillatory components of the task movements, and a second faster range encompassing the information transfer between the interacting individuals.

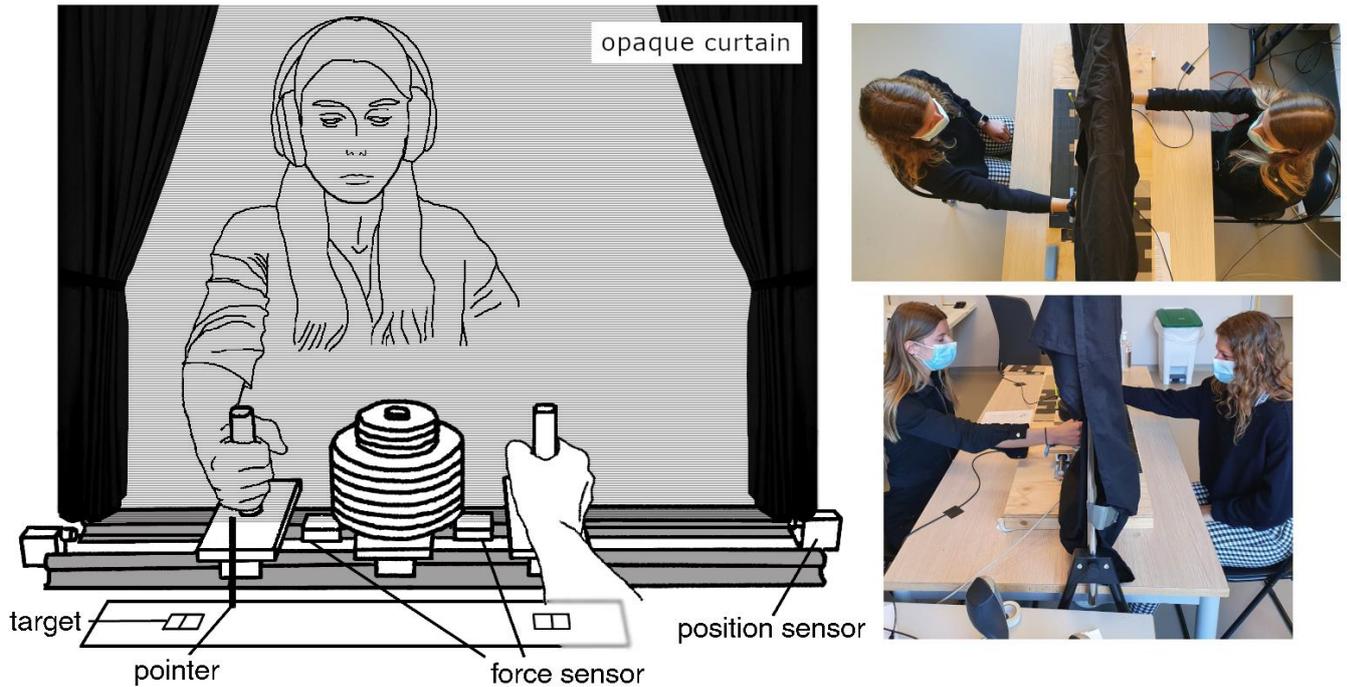

**Fig. 1) Experimental setup.** Our setup consists of a rigid passive slider manipulandum with two handles. The slider slides on two rails on roller bearings to reduce friction. A rack in the center of the manipulandum allows us to load the slider. Position encoders and two 1-dof force sensors (near each handle) allows us to record the participant movements and forces. Participants performed in dyads in our task. They sat on opposite sides of the table and held one handle with their right hands and made reciprocal aiming movement to left and right targets (visible to one participant only in Experiment 1 and to both of them in Experiment 2) while synchronizing as accurately as possible with a metronome provided to them using headphones (one participant only in a dyad in Experiment 1) or speakers (both participants in Experiment 2). A curtain along the axis of the slider prevented the participants from seeing their partner and their handle. The participants were provided with targets toward which they had to perform a reciprocal pointing task with a vertical pointer fixed on the device; they had to aim as accurately as possible at the middle line traced out inside the target box.

## Results

For the sake of clarity, we reintroduce some terminology here that is more developed in the Methods (see Fig. 1 for more information on the task and its setup). In order to increase information exchange between participants, we distributed the type of feedback required for the task completion between them. In each dyad the 'Synch Participant' was given the metronome to be followed (using earphones) but was not provided with the target positions that defined the movement range. The other partner, the 'Target Participant', was provided with the target information but not the metronome (hence the movement timings to be maintained).

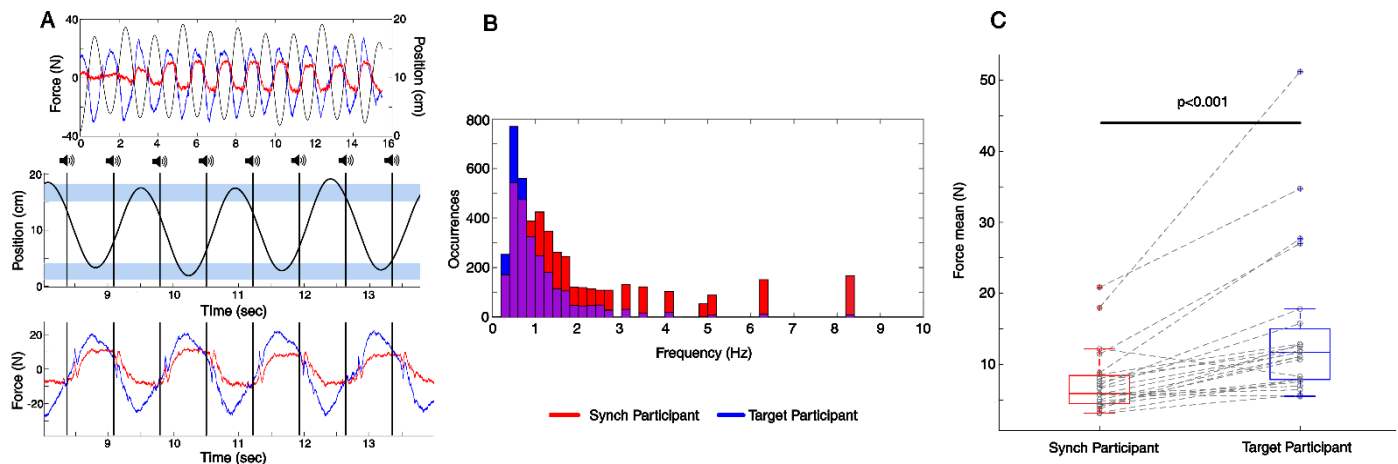

**Fig. 2 Forces time series and task behavior in a representative trial for Experiment 1, histograms of periodic components in forces, and force amplitudes statistics. A)** Example of recorded displacement of the manipulandum (top and middle) and forces (top and bottom) from a representative dyad in Experiment 1. Temporal cues heard by the Synch Participant are represented as vertical black lines, and target area as horizontal pale bands. Participants were required to synchronize their change of direction with the metronome, while aiming at each target as accurately as possible. **B)** Histograms of frequency content of the participant's forces' time series show more higher frequencies in the forces generated by the Synch Participants than the Target Participants. A two-sample Kolmogorov-Smirnov test confirmed the difference between the two distributions (p<0.001, D=0.33). **C)** Average and individuals applied force by participants in each dyad, across Experiment 1. Target Participants applied consistently more force than Synch Participants as confirmed by a Wilcoxon rank sum test (p<0.001, W=385).

Fig. 2A shows an example movement and recorded forces from a representative dyad in one condition. We observed that all dyads could perform the task well even when each individual in the dyad was provided with only one of the required task feedback (targets or metronome). The Target Participants were observed to consistently apply larger force amplitudes than the Synch Participants (p<0.001 W=385, Wilcoxon rank sum test, Fig. 2C). In terms of frequency content though, the Synch Participant's forces were observed to contain a higher frequency content than the Target Participants (p<0.001 D=0.33, two-samples Kolmogorov-Smirnov test, Fig. 2B).

Next, we analyzed the forces collected in the nine conditions using GGC analyses to isolate patterns that were consistent across the different conditions (See Methods for details). Fig. 3A shows the 'GGC values' that provide a quantification at each frequency, of the influence of the forces generated by the Synch Participants on the subsequent forces produced by the Target Participants ($I_{S \to T}$, red trace), as well as the influence of the forces generated by the Target Participants on the subsequent forces produced by the Synch Participants ($I_{T \to S}$, blue trace). The dotted black trace (Fig. 3A) shows the 99th percentiles of the GGC values ($I_{BS}$) obtained using a randomization procedure (bootstrap) of force time series of individuals who were not partners in the task. $I_{BS}$ therefore gives us an estimate of the baseline Granger values that are not related to the interpersonal interaction in the task, but rather result from the non-specific constraints imposed by the task, like its low frequency cyclic nature (seen as the relatively high Bootstrap values in the lower frequencies below 2 Hz), or the tremors in the individual forces (observed as the relatively high Bootstrap values above 9 Hz).

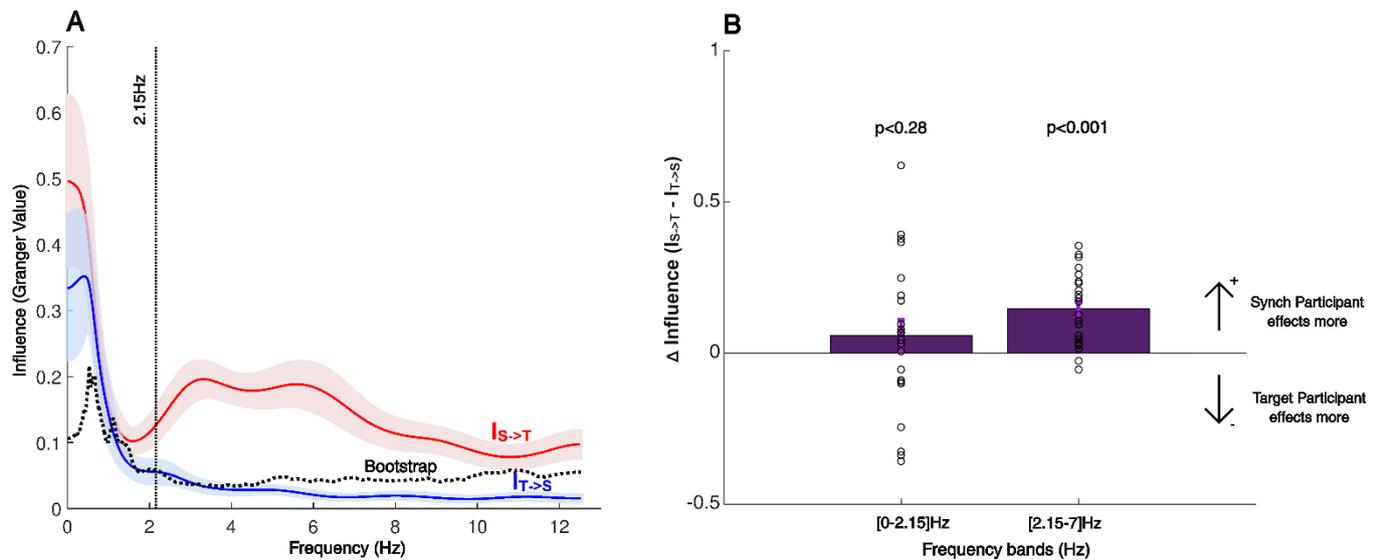

**Fig. 3) Granger-Geweke causality (GGC) spectrum of forces in Experiment 1. A)** Average curve and standard deviation of influence of the Synch Participant on the Target Participant ($I_{S->T}$, red trace) and the Target Participant on the Synch Participant ($I_{T->S}$, blue trace). The average curve and standard deviation were estimated from 46 individuals (23 dyads x 2 directions). A Bootstrap analysis (99th percentiles of null distribution using a bivariate permutation among dyads) was used to identify the baseline Granger values that are independent of interaction. **B)** The overall differences between inter partner influence ($I_{S->T} - I_{T->S}$) are presented for the two frequency of [0-2.15] Hz and [2.15-7] Hz; this Δ influence was the difference between the integral of GGC over the frequency of interest of participants in a dyad, estimated in each of the two bands. Bars represent the mean difference; dots represent the difference for each dyad. The influence of the Synch Participant on the Target Participant ($I_{S->T}$) was similar in the [0-2.15] Hz frequency band (W=174, p<0.28, Wilcoxon signed rank test) but significantly higher in the [2.15-7] Hz frequency band (T(44)=6.39, p<0.001, two-sample t-test). Thorough the article, we used a Shapiro Wilk Test to assess the normality of our data before executing any further statistical analysis and two sample t-test or Wilcoxon rank sum test were used accordingly.

The inter-personal influence ($I_{S->T}$ and/or $I_{T->S}$) was observed to be significantly above the Bootstrap values ($I_{BS}$) at several frequencies between 0 Hz and 7 Hz, which we divided and analyzed over two frequency bands of [0-2.15] Hz and [2.15-7] Hz (choice of the bands explained in the methods section) in which the influence (specifically $I_{S->T}$) was observed to be higher than the bootstrap influence (see Methods for more details). Within these bands we calculated the differences between the integral of GGC values $I_{S->T}$ and $I_{T->S}$ (Fig. 3B).

Due to the repetitive nature of our task, most of the participant force was generated at the end points of the repetitive reaches, in order to decelerate the mobile slider at one target, and then accelerating in the opposite direction toward the next target. Given that the movement frequencies were less than 1 Hz across the dyads in our experiment (see. Fig. 4B), most of the power spectrum of the force therefore corresponded to the movement frequency (see. Fig. 2B). The inter-personal influence ($I_{S->T}$ and/or $I_{T->S}$) was not significantly different in the [0-2.15] Hz band (W=174, p=0.28, Wilcoxon signed rank test, Fig. 3B), which corresponds to the main movement frequencies. There were also no observed differences with a smaller frequency band of [0-1.5] Hz (W=168, p=0.36, Wilcoxon signed rank test).

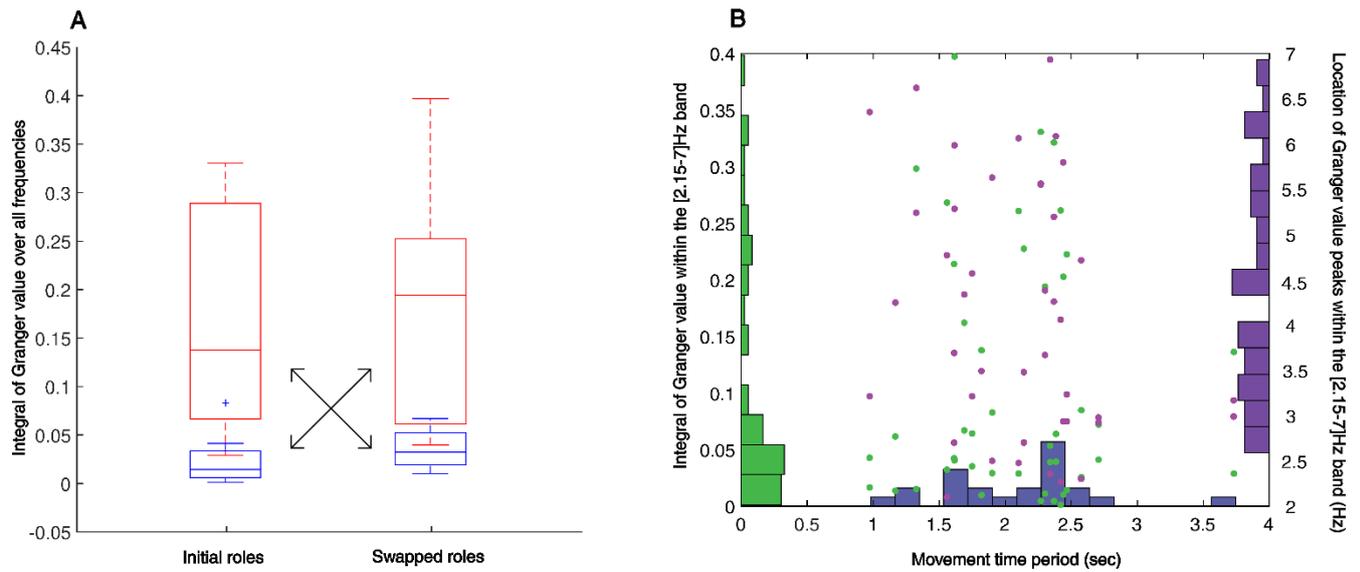

**Fig. 4 Relation between GGC, roles and movement period in Experiment 1. A)** Box plot of the overall influence (Integral of GGC) as a function of roles in the task: Synch (red) vs. Target (blue) medians, 25th and 75th percentiles as well as most extreme data points for 11 dyads are presented in their initial roles and after they swapped roles. Significant differences was found for the same participants when they swapped role from Synch to Target: one-sample t-tests, $p<0.003$, $T(10)=4.05$, and Target to Synch: $p<0.002$, $T(10)=-4.35$). **B)** Histograms of the GGC integrals and GGC peaks in the [2.15-7] Hz band showed no relation between Granger causality and the movement period across participants.

On the other hand we observed that, interestingly, $I_{S \to T}$ was significantly larger than $I_{T \to S}$ ($T(44) = 6.39$, $p<0.001$, two-sample t-test) in the [2.15-7.0] Hz band. This was even though this frequency band was clearly higher than the main movement frequency. Furthermore, across the dyads, by comparing the forces produced by the participants when they swapped their roles, we observed a persistent relation between the roles of the participants (Synch or Target) and the influence exhibited by them (Fig. 4). Significant differences were found when comparing the influence by the same participants when they swapped roles, showing a higher $I_{S \to T}$ compared to $I_{T \to S}$, both for participants who performed the Synch role first, before changing to a Target Participant ($p<0.003$, one sample t-tests, Fig. 4A) or vice versa ($p<0.002$, one sample t-tests, Fig. 4A). No differences were observed when comparing participants adopting the same roles in initials vs. swapped trials ($p=0.997$ and $p=0.187$, Synch and Target roles respectively, one sample t-tests, Fig. 4).

These results indicate that at frequencies higher than 2.15 Hz, the influence by the Synch Participants on the Target Participants dominated the interaction. On the other hand, the influence in the other direction, $I_{T \to S}$ (blue trace in Fig. 3A) remained below the bootstrap value $I_{BS}$. This was so even though, as we have shown before, the Target Participants contributed significantly more in terms of the mean forces in the task (Fig. 2C). But does this asymmetry in influence between the partners at higher frequencies relate to task specific information exchange? To address this issue, we examined the correlation between the Granger values and the task performance by the dyads.

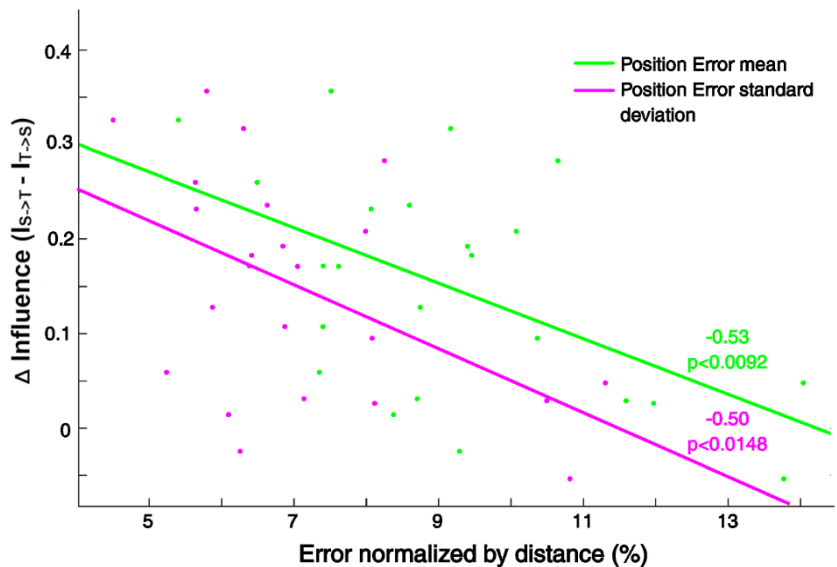

**Fig. 5) Significant correlations between difference of overall influence (integral of GGC values) between participants in the [2.15-7] Hz band and the Position Error performance in Experiment 1.** Negatives correlations of the Position Error mean and standard deviation with the delta Granger Causality in the [2.15-7] Hz frequency band (PEm p<0.009, PEsd p<0.015).

First, we hypothesized that if the influence we isolated is relevant to the task, then the influence should relate to the task's performance. We therefore analyzed the correlation between the mean difference in Influence between the partners within the [2.15-7] Hz band (Fig. 3B) and the target Position Errors and Synchronization Errors exhibited by the dyads. We found a significant correlation between the difference in Influence and the mean (p<0.009, R value -0.53, Pearson correlation) as well as the standard deviation (p<0.015, R value -0.50, Pearson correlation) of the target Position Errors (Fig.5). Dyads with a larger $I_{S \to T}$, compared to $I_{T \to S}$, made lesser mean Position Errors, which also showed lesser standard deviation across time.

Next in Experiment 2, we provided each participant access to sufficient information to achieve the task, attenuating the requirement for interaction. We predicted that this would lead to a change in the pattern of bi-directional influence, and specifically decrease it if the influence found in Experiment 1 did indeed represent information transfer between the dyad partners related to the task.

Experiment 2 followed the same protocol as Experiment 1, except that in Experiment 1, both dyad participants were provided with both types of feedback – the target position as well as the metronome. The dyads again performed in 9 conditions like in Experiment 1, except that this time the metronome time periods were measured as 0.8, 1 or 1.2 times a mean preferred time period, measured in a preliminary experiment with 30 participants who performed the task alone (see Supplementary for a figure showing a representative example of position, forces and frequencies in position in Experiment 2). We observed that unlike in Experiment 1, the forces exerted by the dyad partners were no longer different as a function of roles in Experiment 2 (T(10)=0.88, p=0.40, two-sample t-test).

Figure 6A presents the GGC values in Experiment 2. The influence by each participant on their partner, was not different in either the frequency range [0-2.15] Hz (T(10)=1.25, p=0.24, two-sample t-test) or the frequency range [2.15-7] Hz (T(10)=-1.06, p=0.31, two-sample t-test) (Fig. 6 B) in Experiment 2. Importantly, the overall quantitative influence, measured in a dyad by the sum of partners' integral of GGC values over frequency, was significantly less in Experiment 2 compared to Experiment 1 (T(56)=2.21, p<0.03, two-sampled t-test, Fig. 7). Looking at the performance indices, we also found significant differences in the mean Position Errors and standard deviations between the first and second experiments (PEm: two-sample t-test, p<0.001 T(27)=4.64, and PEsd: Wilcoxon rank sum test, p<0.001 W=412) but no difference between Synchronization error (Wilcoxon rank sum test, SEm: p<0.056 W=381, and SEsd: p<0.11 W=375) (Supplementary figures S3 and S4).

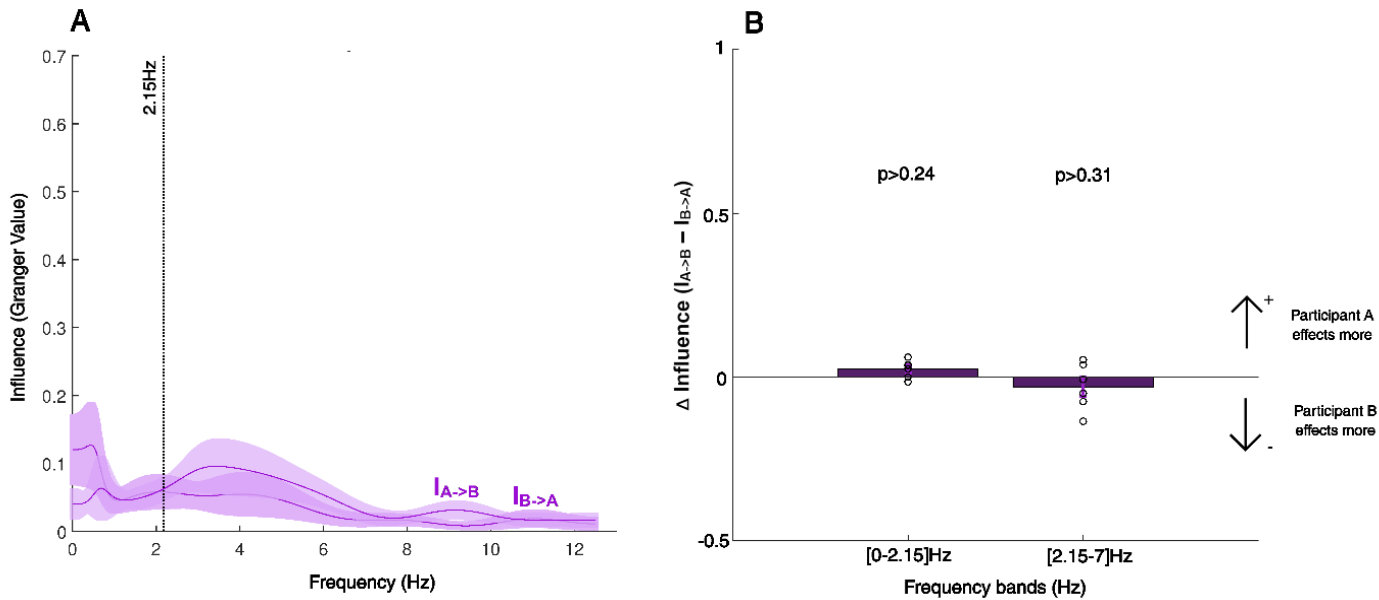

**Fig. 6) Granger-Geweke causality (GGC) spectrum of forces in Experiment 2. A)** GGC values for both participants are shown in purple. Note that given that roles were fully symmetric in Experiment 2 labels A or B were attributed here arbitrarily. The average curve and standard deviation were estimated from 12 individual curves (6 dyads x 2 directions). As in Experiment 1, we analyzed the GGC values (influence) within the two frequency bands: [0-2.15] Hz and [2.15-7] Hz. **B)** Differences in overall influence between participants in a dyad, estimated from integral of GGC values over frequency, in the two frequency bands. No significant differences were found between the two directions of influence using a two-sampled t-test either in the [0-2.15] Hz (p>0.24 T(10)=1.25) of [2.15-7] Hz (p>0.31 T(10)=-1.06) bands.

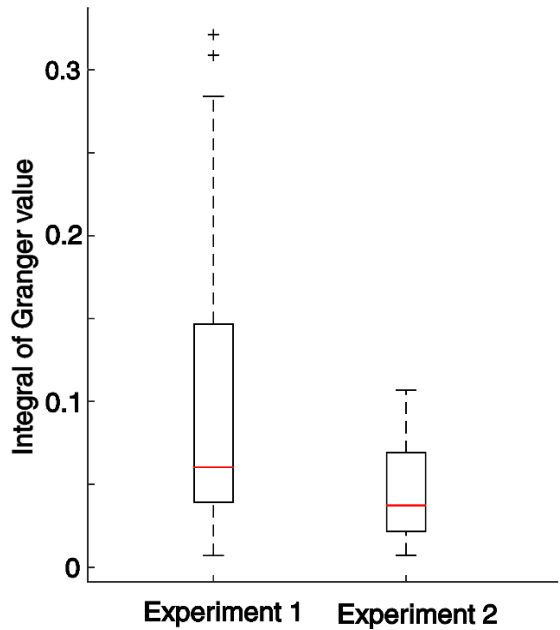

**Fig. 7) Box plot of the total influence, estimated from the integral of GGC values of participants in a dyad across the frequencies, for each experiment.** The total inter-personal influence were significantly less in Experiment 2 compared to Experiment 1 (T(56)=2.21, p<0.03, two-sample t-test). Total inter-personal influence was also significantly less in Bootstrap compared to Experiments 1 and 2 (T(1068)=30.99, p<0.001, two-sample t-test).

**Discussion**

Interpersonal physical interactions are mediated by socials and mechanical factors (5, 6), as well as by multisensory integration of interaction forces with other sensory modalities, like vision or audition. In this study, we focused on the forces generated by the interacting individuals. On the one hand, these forces provide the energy to perform the task. On the other hand, it is believed that these forces are also used to exchange information to infer each other's behaviors and to coordinate in the task (8, 9, 16, 17). In this study, we aimed at isolating the part of the forces that contribute to the inter-personal information exchange. For this, we developed a scenario in which information exchange was imperative for success in completing the task. We chose to do this by limiting the feedback available to the interacting partners (defining them as either Synch- or Target- Participants). We selected this procedure over modulating the quality of (noise in) the feedback available to the partners (9), or limiting their motor ability in a task (16) so as to ensure that both partners are engaged in the task at all times. In our design, complementary roles, defined by distinct perception-action couplings, or type of feedback, instead of redundant roles, enforced partners to cooperate.

We analyzed the forces, generated by the dyad partners during the interaction, using Granger-Geweke causality analysis, expecting a difference in the strength and frequency of influence among in our experiments. We found a significant difference in the influence by the Synch Participants over the Target Participants within the frequency range of [2.15-7] Hz, corroborating our hypothesis. The total influence (Fig. 7) as well as the influence difference between partners (Fig. 6B) within this range, diminished when both partners were provided with enough feedback to perform the task alone in Experiment 2, suggesting that this frequency band served as a channel for information exchange between the interacting individual. This result is further corroborated by the correlations observed between the Granger value differences in this frequency range and the performance errors made by the dyads in the task (Fig. 5).

How much could these results generalize across potential variations of parameters of the tasks and of the movements involved? Here the available knowledge about the frequency content of simple voluntary movements performed by single individuals may offer some indications. Since the seminal studies by Woodworth (1899) (26), a common theme in motor control has been that complete movements are composed of smaller, so-called submovements (29). This particular property of movements is suggested to be general, and hence if submovements are one of the medias of the exchange of information at higher frequency we found, we may anticipate that it generalizes to a wide variety of tasks.

Of particular interest could be the submovements that lie in the range 2-3 Hz, which are often attributed to time-delays in visuo-motor control loops (33). In the case of interaction between individuals, Noy et al. (2011) (30), using a mirror-game in which a follower visually reproduced the leader's movement, found a 2-3 Hz jitter super-imposed to the slower main movement which itself had in average a frequency of 1.5 Hz. In their study, the direction of exchanges was exclusively from the leader to the follower, the former had no way to perceive the movements of the follower. Recently such oscillatory submovements at 2-3 Hz were also found in birectional interaction between two individuals, using a visual synchronization task (33). It is also noteworthy that studies have found neural correlates of submovements frequencies lying between 2 and 5 Hz in primary motor and sensory areas, which were phase-locked to movement speed fluctuating at those frequencies (27). A common 3 Hz oscillation was also recorded in neurons firing in the primary motor cortex leading to 3 Hz submovements in movement kinematics (34). We suggest that the frequency band of [2.15-7] Hz identified in the present study could be related to these various submovements.

Our experiment introduces some differences compared with previous studies discussed above. Furthermore, the weight of the object to be manipulated, and the multisensory processes ascribed to the integration of the haptic sensing of forces could introduce changes in the specific range of frequencies of submovements involved. Despite such foreseeable variations, it is likely that a frequency component higher than the main frequency of the movement is generally involved in mutual exchanges when a dyad cooperates. Further experiments would be required to investigate the generality and boundaries of this mechanism.

A frequency band of [2.15-7] Hz can be argued to be relatively wide. While we observed the inter-personnel influence in this range in all dyads, the frequency at which the influence was most prominent changed across dyads and it remains unclear where there is in fact one key frequency band which dominates the information transfer. One possible candidate may be the physiological delays involved (35). Within the [2.15-7] Hz band we observed that the influence in the [2-4] Hz frequency sub-band correlated better with the task performance (Supplementary Fig. 2). However further studies are required to clarify precisely how information is coded in the forces within the [2.15-7] Hz band.

Granger causality, more precisely Granger Geweke causality (GGC), quantifies the magnitude of causal influence (22, 24) between the partners in an interaction. Its use is also possible to analyze non-repetitive data and non-isochronous sequences. While GGC has been widely used in neurosciences and brain imaging to investigate reciprocal influence in neural networks oscillating at discrete frequencies (25), to the best of our knowledge, this is the first use of GGC to estimate the influence of the forces produced during physical interaction.

In our study, while we apriori expected an increase of the overall bi-directional influence in the first experiment compared to the second, the asymmetric influence between the partners was not expected. We observed that within the [2.15-7] Hz range, the Synch Participants consistently influenced their partner more than the Target Participants, even though the Target Participants consistently produced larger forces and clearly contributed more to the task in terms of the energy expenditure (Fig. 2B). This pattern of cooperation and coupling was spontaneously adopted, without verbal or facial communication, within a short time adaptation window, and persisted after partners switched roles. These results are probably the first to show a dissociation of labor between the partners in terms of the mechanical effort and causal influence.

The present study is not the first to investigate a division of labor amongst its participants. Reed and Peshkin (2008) (38) observed a role division between their participants in a task of rotating a two-handled crank. A subsequent study showed that one participant specialized in producing tangential forces while their partner was involved more in producing radial forces, related to finer positioning (39). A differentiation of roles was also found in a study where two participants had to carry a load together (3), with one participant applying consistently more force than the other, leading the researchers to suggest that the participant applying more force was the leader. On the basis of the present results, it can be argued that even if the Target Participants were the ones producing more force, it was the Synch Participants' that influenced more the other. This suggests that during this interaction the Synch Participants were the leaders and the Target Participants, who were influenced, the followers. Like in Reed and Peshkin (2008) (38), the Synch Participants could be involved in a fine regulation of the movement. It is however noteworthy that an increased influence by the Synch Participant did not improve the Synch performance but instead the target aiming performance. Previous studies have shown that sensorimotor synchronization is accompanied by a decrease of the variability of movement at the times of the metronomes beats (36, 37), aka the anchoring effect. Such spatio-temporal stabilization may in turn improve the dyads' performance in the pointing task. Furthermore, the permutation of roles shows that the asymmetry in GGC between participants depended on their roles during the task and not the individuals performing them (Fig. 4A). This suggests that any cooperation task with division of roles is likely to give rise to such asymmetry of influence. Humans in physical interactions are known to implicitly take a leader or follower role (13) and our results point at the magnitude of causal influence, in the sense of Granger, to offer a proper way to measure the leadership during physical interactions.

It is also noteworthy that small forces have been known to be sufficient to enable communication of movement goals (4) during physical interaction, with exchanges of forces as small as 3N leading to synchronized walking by individuals (18), to 15N forces being deployed during a handshake (40). Therefore, the scale of the forces measured in the present study is about the range of forces previously documented (Fig. 2).

Finally, in the course of the study of a so-called "haptic channel", van der Wel et al. (16) stated that "a complete understanding of how people perform joint actions will require explication of how and when different modalities are used to support performance". It can now be argued that haptic communication takes place through the exchange of relatively small forces in a frequency band greater than the main frequency of movement, in the range [2.15-7] Hz. In future studies, an empirical manipulation of the participants' ability to access those frequencies could be a direct way to put our conclusions to the test.

**Materials and Methods**

Both studies were carried out according to the principles expressed in the Declaration of Helsinki, and were approved by the EuroMov ethical committee (EuroMov IRB #1912A, University of Montpellier). All participants provided their written informed consent to participate in the study, using a consent form reviewed by the ethical committee. In addition, all participants gave their informed consent for publication of identifying images (i.e., Fig.1) in an online open-access publication.

**Experimental Design**

In our study the participants operated a rigid mobile slider with their right hand (See Figure 1). The device is equipped with force and position sensors. An additional mass (13.5kg) is added in its center, making a total weight of 16.5kg for the mobile slider. A curtain separated the set-up in half, preventing the participants from seeing each other. Participants were asked to stay silent during the entirety of the experiment.

We recorded (at 500 Hz) the mobile slider's position using two linear position transducers, and the force applied by each participant using load cells and through 2 A/D cards (NI USB 6229 16-bit Digital Acquisition Board).

48 participants (n=28 in Experiment 1; n=12 in Experiment 2) gave informed consent to participate (age 18 to 40). All participants were right handed and no participant participated to both experiments (♀=12 ♂=16 in Experiment 1; ♀=4 ♂=8 in Experiment 2).

Each participant was weighed at the end of the experiment and the de Leva table (41) was used to calculate their hand and forearm's mass. They also participated in a strength test using a dynamometer (not used in the present analysis). Expert musicians and dancers (10 years of regular practice), as well as people practicing rhythmic or interpersonal coordinative sports were excluded from this study.

**Experiment 1**

The participants performed a cooperative task in which they had to move the mobile slider together repetitively between two targets, and synchronized with a metronome (Figure 1). However, the movement targets and metronome were not provided to both dyad participants. In each dyad, one partner was assigned the role of the "Synch" participant, and was provided with the auditory metronome (through headphones), while the other partner was assigned the role of the "Target" participant, and provided with the targets to guide the task. The auditory metronome was displayed using a custom Matlab® program and PC computer sound card (Intel®), duration = 80ms, sinewave, tone = 500 Hz. Beat events were recorded using an A/D card (NI USB 6002 16-bit Digital Acquisition Board). Both participants were instructed to cooperate to make repeated movements to targets while following a metronome as best as possible, and knew that each participant received only one feedback.

The distance between targets and metronome frequency were varied across 9 conditions. We sued three target distances (close, medium and far distances = 11, 20 and 29cm respectively). The targets' width was 3cm with a line traced out at the center. Three metronomes beat rates were used: One medium beat rate corresponding to the mean preferential frequency of the Target Participant, measured in a preliminary 'solo trial, one 10% slower, and one 10% faster. Targets were changed manually by the experimenter.

One trial lasted 20 beats (between ~16 and ~30 seconds depending on the target condition) and each dyad had to perform three trials per condition. Overall, each dyad performed 27 trials, and the order of the conditions were randomized. Participants then exchanged roles, the Target Participant becoming the Synch Participant and vice versa, and performed a second set of 27 trials.

At the beginning of each trial a random interval of a few seconds of silence was inserted before the beat metronome started, so the participants would not use the experimenter starting instruction as anchor

**Experiment 2**

In this experiment participants were asked to follow the same instructions as in the first experiment. This time the temporal (beats) and spatial (targets) cues were not divided between the participants and both participants could see the targets and hear the beats.

The same 9 conditions as the first experiment, were used in the 2nd experiment. The mean preferential frequency used for the medium speed was the mean preferential frequency of 30 participants for a given distance between targets, then one beat rate was 10% faster and one was 10% slower for the two other metronome conditions. Overall, there were a total of 9 different conditions.

The same auditory metronome was displayed with loudspeakers in this experiment. A random interval was inserted before the beat metronome started at the beginning of each trial. One trial lasted 20 beats (between ~16 and ~30 seconds depending of the target) and each dyad had to do one trial per condition, and conditions were again randomized.

**Data processing**

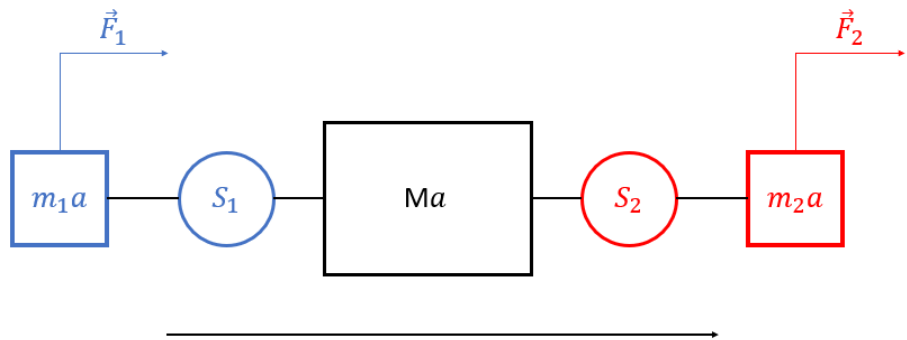

**Fig. 8) Model used to estimate forces from the sensors.** Where $\overrightarrow{m_1 a}$ and $\overrightarrow{m_2 a}$ are the mass (hand + forearm) multiplied by the acceleration, $\overrightarrow{Ma}$ the central mass of the mobile multiplied acceleration, $\overrightarrow{S_1}$ and $\overrightarrow{S_2}$ are the sensor measured force, and $\overrightarrow{F_1}$ and $\overrightarrow{F_2}$ the force applied by the two participants respectively. We consider the effects of friction as negligible.

The mobile slider's position was low pass filtered (Butterworth filter, dual pass, cut off frequency 10 Hz). From the time series of the beats, we detected each onset using the function findpeaks() from Matlab®.
To measure the frequency content of each participant's force time series, we used a local minima and maxima detection method, using the function findpeaks() from Matlab®, to get the distribution of periods, then we converted the periods to frequency in Hz. This method is preferable when time series displays varying amplitudes.

Forces were converted to Newtons and low pass filtered (Butterworth filter, dual pass, cut off frequency 10 Hz). The participant forces were estimated from the sensors according to the model presented in Figure 8.

In Figure 8 $\overrightarrow{m_1 a}$ and $\overrightarrow{m_2 a}$ correspond to the mass (hand + forearm) multiplied by the acceleration, $\overrightarrow{Ma}$ to the central mass of the mobile multiplied by the acceleration, $\overrightarrow{S_1}$ and $\overrightarrow{S_2}$ are the sensor measured forces, and $\overrightarrow{F_1}$ and $\overrightarrow{F_2}$ are the forces applied by the two participants respectively. We considered the effects of friction as being negligible.

Taking a movement toward the right as positive we considered the free body diagram of mass $m_2$ to get:

$$\overrightarrow{F_2} = \overrightarrow{m_2 a} + \overrightarrow{S_2} \qquad (1)$$

Considering the free body diagram of mass $M$:

$$\overrightarrow{S_2} - \overrightarrow{S_1} = Ma \qquad (2)$$

And considering the free body diagram of mass $m_1$ we get:

$$\overrightarrow{F_1} = \overrightarrow{m_1 a} - \overrightarrow{S_1} \qquad (3)$$

Substituting for $a$ from (2) in (1) and (3), we can calculate the force applied by each participant by considering the directions of movements appropriately at any instance, given the mass $M$ (which was fixed in each experiment), and the hand + forearm mass of each participant ($m_1$ and $m_2$), which were estimated for each dyad by weighing each participant and using the de Leva table (41).

In the second experiment, the "roles" of Participant A and Participant B were randomized between the participants of a same dyad, as they were following the same instructions.

**Performance indices**

The Position Error (PE) was calculated using the difference between the position at which our subjects changed direction and the center of the target near the direction change. The Synchronization Error (SE) was calculated as the difference between each period of movement between two targets and the (fixed) period of the metronome for that trial (n = 20). The indices were then expressed in % of the target size (distance) and period of the metronome respectively.

The mean and STD errors were calculated for each of the 9 conditions by every dyad, and then averaged across conditions in order to estimate overall mean and STD errors by one dyad. We found significant differences between the first and second experiment's Position Error means and standard deviations (PEm: two-sample t-test, p<0.001 T(27)=4.64, and PEsd: Wilcoxon rank sum test, p<0.001 W=412). One the other hand, Synchronization Error did not change between the experiments (Wilcoxon rank sum test, SEm : p<0.056 W=381, and SEsd: p<0.11 W=375) (see Supplementary figures S3 and S4). For steady behaviour, the variability estimated by variance or standard deviation is indicative of the (asymptotic) stability against intrinsic continuous stochastic perturbations (42).

The frequency spectrum of the movement of the manipulandum was estimated using Fast Fourier Transform (fft.m from Matlab® Signal processing functions). After visual inspection, a main single peak was identified in the amplitude spectrum and used for subsequent analysis.

For the following steps of the analysis, we excluded dyads whose standard deviation was equal or superior to twice the standard deviation of the errors of the whole sample. We did not exclude any dyad in the second experiment, but in the first experiment our number of dyad (n=14 initially) was reduced at n=11 for the first trial and n=12 for the swapped trial. The high synchronization errors corresponded to a misinterpretation of, or failure to follow, the instructions given by the experimenter.

**Granger-Geweke causality spectral estimation**

Granger-Geweke causality (GCC) was estimated by using the parametric method of estimation (43) from the forces times series. We used the BSmart toolbox in Matlab®, and specifically the functions for bivariate analysis by Cui et al., 2008 (44), for our analysis. Prior to the Granger-Geweke estimation, we down-sampled the forces time series to 25. For each trial and each dyad, the forces time series were segmented into 3 consecutive time windows of equal duration, providing 27 data epochs for each dyad for the GGC analysis. The estimation method provided us the influence from A to B and from B to A, giving for each a vector of Granger values, consisting of one for each frequency bin analyzed. To ensure that the GGC values were not dependent upon the parametric modelling used in the estimation method, we compared it to the non-parametric method (45, 46). We found that both estimation methods produced very comparable GGC values for each dyad. For further analysis we selected the parametric estimation which proved much faster.

According to the Geweke's (24), the Granger causality spectrum from $x_{Bt}$ to $x_{At}$ is computed as follow:

$$I_{B \to A}(f) = -\ln(1 - \frac{\left(\Sigma_{BB} - \frac{\Sigma_{AB}^B}{\Sigma_{AA}}\right)|H_{AB}(f)|^2}{S_{AA}(f)}) \qquad (4)$$

where $\Sigma_{BB}$, $\Sigma_{AB}^B$ and $\Sigma_{AA}$ are elements of the covariance matrix $\Sigma$. $S_{AA}(f)$ is the power spectrum of channel A at frequency $f$ and $H(f)$ is the transfer function of the system (27, 29-31).

**Frequency bands for GGC**

To define two frequency bands of our interest, we first observed that the majority of our dyads had two peaks in the Granger values. We located the first peak for every participant, and calculated the mean and standard deviation across participants. The mean +3*STD across participants was calculated as 2.15 Hz. The first frequency band was thus set between 0.1 Hz and 2.15 Hz. Next, we considered the GGC values for the Synch Participants and noted that the mean +3*STD GGC value went below the bootstrap value at 7 Hz. The second frequency band was thus set between 2.15 Hz and 7 Hz.

To quantify the Granger causality within each band we integrated the GGC values over frequency in each direction of exchange using a trapezoidal numerical integration. Ding et al. (2006) (43) reported the equivalence under general conditions obtained by Geweke (1982) (24) between the Granger causality in the time domain and the integral of the (frequency domain) GGC. Next, for each experiment, we calculated the difference of this integrated GGC between participants in a dyad (Δ influence), for each of the two frequency bands. We used a Wilcoxon signed rank test and a t-test, and found a significative difference only for the first experiment ([2.15-7] Hz p<0.001, T(44)=6.39.

We also analyzed the integral of the GGC values in Experiments 1 and 2 (Fig.7). The total inter-personal influence were significantly less in Experiment 2 compared to Experiment 1 (T(56)=2.21, p<0.03, two-sample t-test). Total inter-personal influence were also significantly less in Bootstrap compared to Experiments 1 and 2 (T(1068)=30.99, p<0.001, two-sample t-test).

**Permutations (bootstrap in the main text) for surrogates GGC spectra**

To examine the influence of individual force time series on the estimation of the GGC spectrum, and confirm the effect of genuine interaction, we used a random shuffling method (25). To test for significant GGC value at each frequency bin, bivariate random permutations were used as a null hypothesis of absence of genuine influence between participants. We combined all conditions and randomly shuffled (n = 506) forces between dyads. The GGC spectra were estimated for the permuted dyads similar to the estimation done for the truly observed dyads. From the distribution of GGC values for the frequency bins analyzed under the null hypothesis, the threshold for significance was taken as the 99 percentiles for each frequency bin (See Figure 3A).

**Main statistical analysis**

We used a Shapiro Wilk Test to assess the normality of our data before executing any further statistical analysis, and two sampled t-test or Wilcoxon rank sum test were used accordingly. To analyze the difference between the applied forces, we performed a Wilcoxon signed rank test on the mean force for each dyad across conditions.

Analyzing the frequency content of the forces, estimated from the periods obtained from local minima and maxima, we found more occurrences of high frequencies in the Synch Participant forces than in the Target Participant forces in Experiment 1. A two-sample Kolmogorov-Smirnov test confirmed the difference between the two distributions, for the Experiment 1 ($p<0.001$, $D=0.33$).

We used Pearson correlation to measure the correlation between the difference of integral of GGC values over frequency ($\Delta$ influence) and our performance indices in Experiment 1. In the [2.15-7] Hz frequency band, we found two significant negative correlations between the $\Delta$ influence and the Position Error mean (PEm) and Position Error standard deviation (PEsd) (PEm $p<0.01$, PEsd $p<0.015$, Fig. 5). To further our analysis, we subdivided both the [0-2.15] Hz and the [2.15-7] Hz frequency bands in smaller frequency bands and examined the same correlations (see Supplementary figures). We found significant correlations in the [2.15-4] and [4-6] Hz frequency bands, with negatives correlations between Position Error performance and $\Delta$ influence ([2.15-4] Hz PEm $p<0.006$, [4-6] Hz PEm $p<0.036$), confirming that the $\Delta$ influence in the [2.5-6] Hz band relates to the performance in our task.

To examine the relation between Granger causality and the variability of performance in the task, we used the standard deviation of the errors in a given trial. The Granger values showed negative correlation with the standard deviation of the Position Error in the [2.15-4] Hz band ($p<0.01$) and [4-6] Hz band ($p<0.043$). Such effects were, however, not seen for the Granger Causality differences in [0-2.15] Hz a [6-7] Hz bands. No significant correlations were observed in Experiment 2.

**Acknowledgments**
Authors thank Ding, Bressler Chen et al. for making the toolbox Bsmart publicly available. We also thank Simon Pla for building the experiment apparatus setup.

**Funding:** This study was financially supported by the European Project ENTIMEMENT, H2020-FETPROACT-2018, Grant Number 824160.

**Competing interests:** Authors declare that they have no competing interests.

**Contributions**
Conceptualization: JL GG CC
Data curation: CC
Formal Analysis: JL GG MD CC
Funding acquisition: JL GG
Methodology: JL GG MD CC
Investigation: CC
Project administration: JL
Resources: JL
Software: JL GG MD CC
Supervision: JL GG
Validation: JL MD
Visualization: GG CC
Writing—original draft: JL GG CC
Writing—review & editing: JL GG MD CC